\documentstyle[12pt,titlepage]{article}

\def\L{{\cal L}}

\def\a{\alpha}

\def\d{\delta}

\def\O{{\cal O}}

\def\gbig {\hbox{\Large\it g}}

\def\k{\kappa}
\def\l{\lambda}
\def\n{\eta}

\def\z{\zeta}

\def\t{\theta}
\def\z{\zeta}

\def\vphi{\varphi}
\def\w{\omega}

\def\hf{\frac{1}{2}}
\def\der{\partial}

\def\bq{\begin{equation}}
\def\eq{\end{equation}}
\def\brr{\begin{eqnarray}}
\def\err{\end{eqnarray}}
\def\ba{\left(\begin{array}}
\def\ea{\end{array}\right)}
\def\pp{\hbox{\ooalign{$\displaystyle\int$\cr$-$}}}

\def\ba{\left(\begin{array}}
\def\ea{\end{array}\right)}
\def\pp{\hbox{\ooalign{$\displaystyle\int$\cr$-$}}}

\def\ppb{\hbox{\ooalign{$-$\cr$\displaystyle\int_{-\infty}^{+\infty}$}}}

\def\gbig {\hbox{\Large\it g}}
\def\wbig {\hbox{\Large\it w}}

\begin{document}
\pagestyle{empty}
\begin{flushright} CERN-TH.7353/94, UPR-621T \end{flushright}
\begin{center}{\bf SUPERSYMMETRIC FIELD THEORY}\end{center}
\begin{center}{\bf FROM}\end{center}
\begin{center}{\bf SUPERMATRIX MODELS}\end{center}
\begin{center} RAM BRUSTEIN$^{*)}$   \end{center}
\vspace{-.3in}
\begin{center} Theory Division, CERN \end{center}
\vspace{-.3in}
\begin{center} CH-1211 Geneva 23, Switzerland  \end{center}

\begin{center} MICHAEL FAUX$^{**)}$ and BURT A. OVRUT$^{**)}$
\end{center}
\vspace{-.3in}
\begin{center} Department of Physics \end{center}
\vspace{-.3in}
\begin{center} University of Pennsylvania \end{center}
\vspace{-.3in}
\begin{center} Philadelphia, Pa 19104-6396, USA \end{center}
\begin{center} ABSTRACT \end{center}

We show that  the continuum limit of one-dimensional
${\cal N}=2$ supersymmetric matrix models can be described by a
two-dimensional  interacting  field theory of a massless
boson and two chiral fermions.  We interpret this field
theory as  a two-dimensional ${\cal N}=1$ supersymmetric
theory of two chiral superfields, in which one of the chiral
superfields has a non-trivial  vacuum expectation value.

\vspace*{1.8cm}
\noindent
\rule[.1in]{16.5cm}{.002in}\\
\noindent
$^{*)}$\ \ Contribution to the Proceedings of  the IV International Conference
on Mathematical \\ \indent Physics,
 String Theory and Quantum Gravity, Rakhov, Ukraine, 12-20 February 1994.\\
$^{**)}$\ \ Work supported in part by DOE under Contract No.
DOE-AC02-76-ERO-3071.
\vspace*{0.5cm}

\begin{flushleft} CERN-TH.7353/94 \\
July 1994
\end{flushleft}
\vfill\eject

\pagestyle{plain}
\setcounter{page}{1}

To use $d=1$ matrix models \cite{done,marpar} for the purpose of understanding
non-perturbative effects in superstring theory, it is essential to first
construct the complete two-dimensional effective Lagrangian for the associated
$d=2$ superstring theory.
We present  the effective Lagrangian and discuss its properties. Our
presentation  is based on \cite{sbfo}.
More details and explicit calculations are included there.

A class of $d=1,\ {\cal N}=2$ supermatrix models may be defined using a matrix
superfield,
\bq \Phi_{ij}=M_{ij}(t)+i\t_1\Psi_{1ij}(t)+i\t_2\Psi_{2ij}
    +i\t_1\t_2F_{ij}(t),
 \eq
where $\t_1$ and $\t_2$ are real anticommuting parameters,
$M_{ij}$ and $F_{ij}$ are $N\times N$ bosonic Hermitian matrices and
$\Psi_{1ij}$ and $\Psi_{2ij}$ are $N\times N$
fermionic Hermitian matrices.
A manifestly invariant  Lagrangian,
\bq L=\int d\t_1d\t_2\Biggl\{\hf Tr D_1\Phi D_2\Phi+iW(\Phi)\Biggr\},
 \label{slag}\eq
can be written, using $\Phi$ and the
covariant derivatives
\bq D_I=\frac{\der}{\der\t_I}+i\t_I\frac{\der}{\der t},\ \ \
I=1,2.\eq
The superpotential $W$ is  a real polynomial in $\Phi$,
\bq W(\Phi)=\sum_n b_n Tr\Phi^n.
\eq
In terms of the component functions, Lagrangian (\ref{slag}) is the following
  \brr L &=& \sum_{ij}\Biggl\{\hf(\dot{M}_{ij}\dot{M}_{ji}+F_{ij}F_{ji})
     +\frac{\der W(M)}{\der M_{ij}}F_{ij}\Biggr\} \nonumber \\
     & & -\frac{i}{2}\sum_{ij} (\Psi_{1ij}\dot{\Psi}_{1ji}
     +\Psi_{2ij}\dot{\Psi}_{2ji})
     -i\sum_{ijkl}\Psi_{1ij}\frac{\der^2W(M)}{\der M_{ij}\der M_{kl}}
     \Psi_{2kl}.
 \label{lagmatsusy} \err
The supersymmetry transformations of the component functions are
\brr \d M_{ij} &=& i\n^1\Psi_{1ij}+i\n^2\Psi_{2ij} \nonumber \\
     \d\Psi_{1ij} &=& \n^1\dot{M}_{ij}+\n^2F_{ij} \nonumber \\
     \d\Psi_{2ij} &=& \n^2\dot{M}_{ij}-\n^1F_{ij} \nonumber \\
     \d F_{ij} &=& i\n^2\dot{\Psi}_{1ij}-i\n^1\dot{\Psi}_{2ij},
 \err
where $\n^1$ and $\n^2$ are anticommuting constants.

The classical theory possesses, in addition to  supersymmetry, a global
$U(N)$ symmetry. To verify this fact note that $\Phi_{ij}$ remains a Hermitian
matrix of superfields under the transformation
$\Phi\rightarrow{\cal{U}}^\dagger\Phi{\cal{U}}$,
 where ${\cal{U}}$ is an arbitrary $N\times N$ matrix of complex numbers. The
Lagrangian is invariant under such a transformation, provided that
${\cal{U}}\in
U(N)$.

We restrict our attention to the sector of the theory that is a
singlet under the global $U(N)$ symmetry.  After eliminating the auxiliary
fields using their equations of motion,
the singlet sector can be described in terms of the eigenvalues $\l_i$ of the
bosonic matrix $M$, and their fermionic superpartners $\chi_{i}$. $\chi_{i}$
are the diagonal elements  of the matrix $\chi=U\Psi U^\dagger$, where $U$ is
the matrix used to diagonalize $M$. Note that $U$ diagonalizes $M$, but that
$\chi$ is not diagonal. The Lagrangian for the singlet sector is given by
\cite{dabh,sbfo}
\brr L &=& \sum_i\Biggl\{\hf\dot{\l}_i^2
     -\hf\left(\frac{\der W}{\der\l_i}\right)^2
     -\frac{\der \wbig}{\der\l_i}\frac{\der W}{\der\l_i}
     -\hf\left(\frac{\der \wbig}{\der\l_i}\right)^2
     -\frac{i}{2}(\bar{\chi}_i\dot{\chi}_i
     -\dot{\bar{\chi}}_i\chi_i)\Biggr\} \nonumber \\
     & & -\sum_{ij}
     \Biggl\{\frac{\der^2 W}{\der\l_i\der\l_j}\bar{\chi}_i\chi_j
     +\frac{\der^2 \wbig}{\der\l_i\der\l_j}\bar{\chi}_i\chi_j\Biggr\}.
 \label{lageig} \err
The induced superpotential,
\bq \wbig=-\sum_{j\ne i}\ln|\l_i-\l_j|,
 \label{dasher} \eq
represents a repulsive interaction between the bosonic eigenvalues.

In preparation for taking the continuum limit, it is useful to
change variables, thus defining three collective fields,
\brr \vphi(x,t) &=& \sum_i\Theta(x-\l_i(t)) \nonumber \\
     \psi(x,t) &=& -\sum_i\d(x-\l_i(t))\chi_{i}(t) \nonumber \\
     \bar{\psi}(x,t) &=& -\sum_i\d(x-\l_i(t))\bar{\chi}_{i}(t).
 \label{colldef} \err
Note that (\ref{colldef}) is nothing but a  change of variables.
It does not increase or decrease the number of dynamical variables. In terms of
the collective fields,
\brr L &=& \int dx\Biggl\{
       \frac{\dot{\vphi}^2}{2\vphi'}
       -\hf\vphi'W'(x)^2
       +\frac{W''(x)}{\vphi'}\bar{\psi}\psi \nonumber \\
       & & -\frac{1}{2\vphi'}(\bar{\psi}\dot{\bar{\psi}}
       +\dot{\bar{\psi}}\bar{\psi})
       +\frac{i}{2}\frac{\dot{\vphi}}{\vphi^{'2}}
       (\bar{\psi}\psi'-\bar{\psi}'\psi)\Biggr\} \nonumber \\
       & & +\frac{1}{3}\pp dxdydz\frac{\vphi'(x)\vphi'(y)\vphi'(z)}
       {(x-y)(x-z)}, \nonumber \\
       & & +\pp dxdy\frac{\vphi'(x)\vphi'(y)}{(x-y)}W'(x)
       \nonumber \\
       & & +\pp\frac{1}{(x-y)}\Biggl\{\bar{\psi}(x)\psi'(y)
       -\frac{\vphi''(y)}{\vphi'(x)}\bar{\psi}(x)\psi(x)\Biggr\}.
\label{lagcollpp} \err
The measure of the path integral has, of course, to be changed accordingly. We
will not do that explicitly here, since we will be interested in regions of
parameter space in which the measure takes a  simple form.

 The continuum limit of bosonic matrix models is known to be a
two-dimensional  field theory \cite{dj,jevrev}. We expect, therefore,
that the continuum limit of supermatrix models
is a two-dimensional field theory as well. However, we will see that the
number of fields that survive in the continuum limit is larger in the
supersymmetric case.  Taking the continuum limit consists of a few separate and
independent steps which supply
the original  supermatrix models with additional information and
should be considered as part the definition of  the theory.  At each step  some
choices have to be made, each determines essential properties of the resulting
models.  It is at this juncture that
the field content and specific background are chosen.
Previously, some attempts were made, with varying degree of success, to obtain
the correct two-dimensional continuum field theory [4, 7--10].
The first step, necessary to ensure that the number of dynamical
variables is enough to describe  a two-dimensional field theory is simply \bq
N\rightarrow\infty.\eq
It is useful to think about $N$ as the cutoff, in momentum space, of the
theory. Then a regularization procedure has to be chosen to ensure that all
terms in the Lagrangian are finite as the cutoff is taken to infinity. We use
\bq \ppb dx\frac{\phi(x)}{x-a}=\pm i\pi\phi(a)
 \label{equu} \eq to define our regularization scheme. In addition, some
dependence of the coupling parameters in the superpotential $W$,  on the
cutoff $N$, has to be chosen, placing the theory within a specific
universality class. On general grounds, $W=N W(\frac{x}{\sqrt{N}})$.
Our choice is the following
\brr
W(x)&=&\sqrt{N}c_1x+\frac{1}{6}\frac{c_3}{\sqrt{N}}x^3+\cdots\nonumber \\
&& c_1 c_3<0.
\label{potform}\err
The terms denoted by $\cdots$ are of  higher power in $x/\sqrt{N}$ and do not
change the universality class for non-vanishing $c_i,\ i=1,3$.
For completeness we list some useful expressions for the derivatives of $W$
\brr W'(x) &=& \sqrt{N}c_1+\hf\frac{c_3}{\sqrt{N}}x^2
     +\cdots, \nonumber \\
     W'(x)^2 &=& Nc_1^2+c_1c_3x^2+\frac{1}{3}\frac{c_1c_4}{\sqrt{N}}x^3
     +\cdots, \nonumber \\
     W''(x) &=& \frac{c_3}{\sqrt{N}}x+\cdots.
 \label{drone} \err
The result of  applying  all the steps above to the Lagrangian
(\ref{lagcollpp}), taking (\ref{drone}) into account,
 is  the continuum Lagrangian
\brr L &=& \int dx\Biggl\{\frac{\dot{\vphi}^2}{2\vphi'}
     \pm\frac{\pi^2}{6}\vphi^{'3}
     +\hf\w^2x^2\vphi' \nonumber \\
     & & \hspace{.3in}
     -\frac{i}{2\vphi'}(\psi_1\dot{\psi}_1+\psi_2\dot{\psi}_2)
     \pm\frac{i\pi}{2}\psi_1\psi_1'
     \pm\frac{i\pi}{2}\psi_2\psi_2' \nonumber \\
     & & \hspace{.3in}
     +\frac{i}{2}\frac{\dot{\vphi}}{\vphi^{'2}}
     (\psi_1\psi_1'+\psi_2\psi_2')\Biggr\}.
\label{lagie}\err
Note that there are still three  ambiguous signs in the previous Lagrangian,
related to the sign ambiguity in (\ref{equu}). For the first sign we choose a
minus sign, corresponding to Minkowski spacetime.
Our choice is a minus sign for the second and a plus sign for the third. This
choice determines the chiralities of the
fermions.

A classical static solution of the equations of motion derived
from (\ref{lagie}) is given by
\brr {\psi}_{10}&=&0 \nonumber\\
{\psi}_{20}&=&0   \nonumber\\
{\vphi}_0'&=&\frac{1}{\pi}\sqrt{\w^2x^2-1/{ g}}. \err
We expand around that classical solution
\brr  \vphi&=&{\vphi}_0(x)+\frac{1}{\sqrt{\pi}}\z\nonumber\\
\psi_+ &=& \frac{2^{1/4}}{\sqrt{\pi}}\psi_1 \nonumber \\
     \psi_- &=& \frac{2^{1/4}}{\sqrt{\pi}}\psi_2
  \err
and change coordinates,
\brr \tau'(x) &=& \frac{1}{\pi}
     ({\vphi}_0'(x))^{-1} \nonumber \\
     &=& \frac{1}{\sqrt{\w^2x^2-1/g}}, \err
to obtain
\brr L &=& \int d\tau\Biggl\{
       \hf(\dot{\z}^2-\z^{'2})
       -\frac{i}{\sqrt{2}}(\psi_+\dot{\psi}_+-\psi_+\psi_+')
       -\frac{i}{\sqrt{2}}(\psi_-\dot{\psi}_-+\psi_-\psi_-')
       \nonumber \\
       & & -\hf\frac{\gbig(\tau)\dot{\z}^2\z'}{1+\gbig(\tau)\z'}
       -\frac{1}{6}\gbig(\tau)\z^{'3} \nonumber \\
       & &
       +\frac{i}{\sqrt{2}}\frac{\gbig(\tau)\z'}{1+\gbig(\tau)\z'}
       (\psi_+\dot{\psi}_++\psi_-\dot{\psi}_-) \nonumber \\
       & & +\frac{i}{\sqrt{2}}\frac{\gbig(\tau)\dot{\z}}
       {(1+\gbig(\tau)\z')^2}(\psi_+\psi_+'+\psi_-\psi_-')\Biggr\}
       +\frac{1}{3}\int d\tau\frac{1}{\gbig(\tau)^2}. \label{lagcollfinal} \err
The coupling parameter of the theory varies in space
\bq \gbig(\tau)=
     4\sqrt{\pi}g\frac{\frac{1}{\k} e^{- 2\w(\tau-\tau_0)}}
     {(1-\frac{1}{\k}e^{- 2\w(\tau-\tau_0)})^2}.
 \label{gdef}\eq
We are now in a position to take stock of the field content
of the theory.  Looking at the quadratic terms in the first line of
Eq.(\ref{lagcollfinal}), we observe that the theory contains   one  massless
bosonic field $\z$, and two chiral massless fermions $\psi_\pm$. The
chiralities of the fermions are determined by the choice of signs in
(\ref{lagie}).  If we choose them as we did they have opposite chiralities and
the field content can be fitted within a chiral superfield of a $(1,1)$
two-dimensional supersymmetry.

The Lagrangian  (\ref {lagcollfinal}) is not supersymmetric. It is not even
Poincare invariant.  Motivated by the expected relation to string theory, and
based on our experience in interpreting
the bosonic theory \cite{bo,bda}, we interpret it as follows. We assume that
the theory really started out as a two-dimensional
supersymmetric theory, containing two superfields, $\Phi_1$
and $\Phi_2$. The superfield $\Phi_2$ obtains a non-trivial
vacuum expectation value (VEV). The VEV breaks Poincare invariance as well as
supersymmetry. Our task then becomes
to reconstruct the original theory as best as we can. As
will become obvious, it is not possible to reconstruct the
theory completely. We can, however, capture enough of its
features to make the reconstruction an interesting enterprise.

Of the two chiral superfields of $(1,1)$ supersymmetry,
\brr
\Phi_1&=&\z+i\t^+\psi_++i\t^-\psi_-+i\t^+\t^-Z \nonumber \\
\Phi_2&=&\a+i\t^+\chi_++i\t^-\chi_-+i\t^+\t^-A, \err
it is $\Phi_1$ that contains the degrees of freedom in the original Lagrangian.
It is straightforward to write a manifestly invariant kinetic term for
$\Phi_1$,
\brr \L_{01}^{(eff)} &=& \int d\t^+d\t^-D_+\Phi_1D_-\Phi_1 \nonumber \\
        &=& \hf(\dot{\z}^2-\z^{'2})-i\psi_+\der_-\psi_+
         -i\psi_-\der_+\psi_-+Z^2.
\label{lago1} \err
Doing the same for the other superfield is a little bit more involved
procedure. We expect the superfield $\Phi_2$ to acquire a non-trivial VEV.
Based on our experience in the interpretation of the bosonic theory and
motivated by the expected relation with other formulations of two-dimensional
superstring theory, we expect the components of $\Phi_2$ to obtain the
following VEV
\brr  <\a>       &=& e^{-\w|\tau-\tau_0|} \nonumber \\
      <\chi_\pm> &=& 0 \nonumber\\
      <A>        &=& 0.
\label{moon} \err
Furthermore, we impose that the most singular term in the Lagrangian has a
$1/\a^4$ dependence, in agreement with  classical string theory. We now desire
a manifestly supersymmetric Lagrangian that $(i)$ has (\ref{moon})  as a
solution to its equations of motion, $(ii)$  when the VEV, (\ref{moon}), of
$\Phi_2$ is substituted into the Lagrangian, it reduces to the constant term in
(\ref{lagcollfinal}), and $(iii)$  the most singular term in the Lagrangian has
a $1/\a^4$ dependence. A solution with the desired properties is given by
\brr \L_{02}^{(eff)}&=&\int d\t^+d\t^-\Biggl\{
    F_1(\Phi_2)D_+\Phi_2D_-\Phi_2
    -\frac{1}{\w^2}F_2(\Phi_2)\der_-D_+\Phi_2\der_+D_-\Phi_2\Biggr\}  \nonumber
\\  &=& F_1(\a)\der_+\a\der_-\a-\frac{1}{\w^2}F_2(\a)(\der_+\der_-\a)^2
    \nonumber \\
    & & +F_1(\a)A^2-\frac{1}{\w^2}F_2(\a)\der_+A\der_-A \nonumber \\
    & & -iF_1(\a)\chi_+\der_-\chi_+
        -\frac{i}{\w^2}F_2(\a)\der_-\chi_+\der_+\der_-\chi_+
        \nonumber \\
    & & -iF_1(\a)\chi_-\der_+\chi_-
        -\frac{i}{\w^2}F_2(\a)\der_+\chi_-\der_-\der_+\chi_-,
\label{lo2}\err
where
\brr F_1(\Phi_2) &=& -\frac{1}{48\pi\k\w^2 g^2}
     \left(\frac{11}{5}\frac{\k^3}{\Phi_2^6}
     -\frac{28}{3}\frac{\k^2}{\Phi_2^4}
     +18\frac{\k}{\Phi_2^2}
     -4
     +\frac{5}{3}\frac{\Phi_2^2}{\k}\right) \nonumber \\
     F_2(\Phi_2) &=&  -\frac{1}{48\pi\k\w^2 g^2}
     \left(-\frac{2}{5}\frac{\k^3}{\Phi_2^6}
     +\frac{8}{3}\frac{\k^2}{\Phi_2^4}
     -12\frac{\k}{\Phi_2^2}
     -8
     +\frac{2}{3}\frac{\Phi_2^2}{\k}\right).
 \err

This Lagrangian indeed has (\ref{moon}) as  a solution of its equation of
motion.  Obviously, (\ref{lo2}) is not the unique Lagrangian with the desired
properties. However, it is the Lagrangian with the least number of terms. We
therefore choose to present it. The fact that we were able to find any solution
to our requirements is not at all trivial.

So far, we were able to construct a manifestly supersymmetric theory, to lowest
order in the coupling $\gbig(\tau)$,  using the two superfields and their
covariant derivatives.  Amazingly enough, there exist a manifestly
supersymmetric  Lagrangian that reduces to the full non-linear interacting
two-dimensional field theory. The details of the derivation are given in
\cite{sbfo}. We give the final result here,
\brr &&\L^{(eff)} =
     \int d\t_+d\t_-\Biggl\{
     D_+\Phi_1D_-\Phi_1  \nonumber \\
     & & \hspace{.6in}
     +F_1(\Phi_2)D_+\Phi_2D_-\Phi_2
    -\frac{1}{\w^2}F_2(\Phi_2)\der_-D_+\Phi_2\der_+D_-\Phi_2 \nonumber \\
     & & \hspace{.6in}
     -\frac{f(\Phi_2)}{\w^3\Phi_2^3}
     \frac{\der_{(+}\Phi_1\der_{-)}\Phi_2
           \der_{[+}\Phi_1\der_{-]}\Phi_2}
          {1+\frac{f(\Phi_2)}{\w\Phi_2}
           \der_{(+}\Phi_1\der_{-)}\Phi_2}
           D_{(+}\Phi_1D_{-)}\Phi_2 \nonumber \\
     & & \hspace{.6in}
     +\frac{1}{3}\frac{f(\Phi_2)}{\w^5\Phi_2^5}
     (\der_{[+}\Phi_1\der_{-]}\Phi_2)^3
     D_+\Phi_2D_-\Phi_2 \nonumber \\
     & & \hspace{.6in}
     -\frac{f(\Phi_2)}{\w^5\Phi_2^5}
     \frac{(\der_{[+}\Phi_1\der_{-]}\Phi_2)^2
     \der_{(+}\Phi_1\der_{-)}\Phi_2}
     {1+\frac{f(\Phi_2)}{\w\Phi_2}
     \der_{(+}\Phi_1\der_{-)}\Phi_2}
     D_+\Phi_2D_-\Phi_2 \Biggr\},
 \label{efflag} \err
where
\brr      f(\Phi_2) &=& 4\sqrt{\pi}g\frac{\frac{1}{\k}\Phi_2^2}
         {(1-\frac{1}{\k}\Phi_2^2)^2}.
 \err
In components, (\ref{efflag}) is given by
\brr \L^{(eff)} &=& +\der_+\z\der_-\z
     +Z^2
     -i\psi_+\der_-\psi_+
     -i\psi_-\der_+\psi_- \nonumber\\
    & & +F_1(\a)\der_+\a\der_-\a
         -\frac{1}{\w^2}F_2(\a)(\der_+\der_-\a)^2
    \nonumber \\
    & & +F_1(\a)A^2-\frac{1}{\w^2}F_2(\a)\der_+A\der_-A \nonumber \\
    & & -iF_1(\a)\chi_+\der_-\chi_+
        -\frac{i}{\w^2}F_2(\a)\der_-\chi_+\der_+\der_-\chi_+
        \nonumber \\
    & & -iF_1(\a)\chi_-\der_+\chi_-
        -\frac{i}{\w^2}F_2(\a)\der_+\chi_-\der_-\der_+\chi_-
        \nonumber \\
    & & +\sum_n{\cal{O}}(\a^n\chi_+\chi_-+\a^{n-1}A\chi_+\chi_-).
     \nonumber \\
     & & -\hf\frac{f(\a)\dot{\z}^2\z'}{1+f(\a)\z'}
     -\frac{1}{6}f(\a)\z^{'3}\nonumber \\
     & & +\frac{i}{\sqrt{2}}\frac{f(\a)\z'}{1+f(\a)\z'}
     (\psi_+\dot{\psi}_++\psi_-\dot{\psi}_-)
     +\frac{i}{\sqrt{2}}
     \frac{f(\a)\dot{\z}}{[1+f(\a)\z']^2}
     (\psi_+\psi_+^{'}+\psi_-\psi_-^{'})
     \nonumber \\
     & &
     +\O\Biggl\{\der\z(\psi\chi+\chi\chi+Z\psi\chi+Z\chi\chi
          +A\psi\psi+A\psi\chi)+\psi\psi\chi+\psi\chi\chi\Biggr\}.
  \err
As can be checked, the general solution of the equations of motion derived from
(\ref{efflag}) is the following,
\brr <\a> &=& \exp\Biggl\{
     \w[|t-t_0|\sinh\t_0-|\tau-\tau_0|\cosh\t_0]\Biggr\}
     \nonumber \\
     <\z> &=& constant \nonumber \\
     <\chi_\pm> &=& \n_0^\pm<\a>  \nonumber \\
     <\psi_\pm> &=& 0 \nonumber \\
     <A> &=& 0 \nonumber \\
     <Z> &=& 0.
 \label{solongen} \err
If  this solution is substituted back into (\ref{efflag}) and the auxiliary
fields are eliminated through their equations of motion,
the result exactly reproduces (\ref{lagcollfinal}).

The Lagrangian (\ref{efflag}) has some interesting properties.
First, the superfield $\Phi_1$ has only derivative interactions,
and so, in particular,  has no superpotential. The interactions of the
superfield $\Phi_2$ always contain some derivatives, therefore
the superfield $\Phi_2$ has no superpotential as well. The coupling parameter
of the theory is field-dependent. This is a typical situation in low-energy
effective field theories of string theory. The overall coupling strength is
determined by the parameter $g$, which is sometimes called the ``string
coupling constant". However, if $\Phi_2$ has a space-dependent VEV, as in
(\ref{moon}), the coupling strength varies in space-time and even blows up at
some finite point, signalling the possible existence of new physical phenomena.

\end{document}